\documentclass[aps, twocolumn, nofootinbib,floatfix, showpacs, preprintnumbers]{revtex4} %eqsecnum,
\usepackage{graphicx}
\input epsf

\def\cmm2{{\,\rm cm^{-2}}}
\def\cm2{{\,{\rm cm}^2}}
\def\cmm3{{\,{\rm cm}^{-3}}}
\def\gcmm3{{\,{\rm g\,cm^{-3}}}}

\def\fun#1#2{\lower3.6pt\vbox{\baselineskip0pt\lineskip.9pt
  \ialign{$\mathsurround=0pt#1\hfil##\hfil$\crcr#2\crcr\sim\crcr}}}

\def\be{\begin{equation}}
\def\ee{\end{equation}}
\def\bea{\begin{eqnarray}}
\def\eea{\end{eqnarray}}

\newcommand{\ec}[1]{equation~(\ref{eq:#1})}

\newcommand{\eql}[1]{\label{eq:#1}}

\newcommand{\dd}{{\rm d}}
\newcommand\rp{r_p}
\begin{document}
%\baselineskip=24pt
%\twocolumn[\hsize\textwidth\columnwidth\hsize\csname @twocolumnfalse\endcsname
%\pagestyle{empty}
%\begin{center}
%\rightline{{\large DRAFT} (Ewan; July 28, 1999)}
%\bigskip
%\rightline{FERMILAB--Pub--000-0000}
\preprint{FERMILAB-PUB-07-045-A}
%\rightline{astro-ph/0002360}
%\rightline{submitted to {\it Phys. Rev. Lett.}}

%\vspace{.2in}
\title{Weak Lensing of Baryon Acoustic Oscillations}

 \author{Alberto Vallinotto}
 \email{vallinot@iap.fr}
 \affiliation{Institut d'Astrophysique de Paris,
 CNRS-UMR 7095, Universit\'{e} Paris VI Pierre et Marie Curie,
 98 bis boulevard Arago, 75014 Paris, France,}

\author{Scott Dodelson}
 \email{dodelson@canis.fnal.gov}
 \affiliation{Center for Particle Astrophysics, Fermi National Accelerator Laboratory,
 Batavia, IL~~60510-0500, USA,\\
 Department of Astronomy \& Astrophysics,~The University of Chicago, Chicago, IL~~60637-1433, USA,}

 \author{Carlo Schimd}
 \email{carlo.schimd@cea.fr}
 \affiliation{DAPNIA, CEA Saclay, 91191 Gif-sur-Yvette cedex, France,}

 \author{Jean-Philippe Uzan}
 \email{uzan@iap.fr}
 \affiliation{Institut d'Astrophysique de Paris,
 CNRS-UMR 7095, Universit\'e Paris VI Pierre et Marie Curie,
 98 bis boulevard Arago, 75014 Paris, France.}

\date{\today}
%\smallskip
\begin{abstract}
Baryon Acoustic Oscillations (BAO) have recently been observed in
the distribution of distant galaxies. The height and location of the
BAO peak are strong discriminators of cosmological parameters. Here
we consider the ways in which weak gravitational lensing distorts
the BAO signal. We find two effects that can affect the height
of the BAO peak in the correlation function at the percent level but
that do not significantly impact the position of the peak and the
measurement of the sound horizon. BAO turn out to be robust
cosmological standard rulers.
\end{abstract}


\maketitle

\section{Introduction}

In the early universe, before the recombination of protons and
electrons into neutral hydrogen, the photons, electrons, and protons
were tightly coupled and behaved like a single fluid. The density of
this fluid underwent acoustic oscillations until recombination.
These oscillations left their imprints in the spectrum of
anisotropies in the cosmic microwave background (CMB) as well as in
the distribution of matter via the correlation function of the
galaxy distribution. The effect in the former case is quite
pronounced because the photons have essentially traveled freely
since their last scattering at recombination. In the latter case,
called {\it baryon acoustic oscillations} (BAO), the effect is
diluted since the dominant dark matter did not participate in the
primordial dance~\cite{meiksin-1998-304,Seo:2003pu,Blake:2003rh,Hu:2003ti}.
Observations of both of these effects over the last few
years~\cite{Spergel:2006hy,Eisenstein:2005su} provide dramatic proof
that our picture of the early universe is consistent.

We are now free to go further and use these observations to
constrain cosmological parameters. In the case of BAO, the
correlation function should peak at a comoving scale of order
$\rp\simeq 100 h^{-1}$Mpc where the Hubble constant is parameterized
as $H_0=100h$ km sec$^{-1}$ Mpc$^{-1}$. In a survey of galaxies at
the same redshift, the peak shows up at an angular scale
$\delta\theta=\rp/[(1+z)D_A(z)]$ where $D_A(z)$ is the angular
diameter distance out to redshift $z$. In a 1D radial survey with
redshifts, the peak will show up at $\delta z = \rp H(z)$. More
generally, in a 3D survey, one can hope to measure both $D_A(z)$ and
$H(z)$. These quantities are
enormously important to cosmologists because they depend on the
energy density back to redshift $z$ and therefore on the properties
of matter and dark energy~\cite{Copeland:2006wr,Uzan:2006mf,
Peacock:2006kj,Albrecht:2006um}.
Besides its location, the height of the BAO peak is governed by the
matter density
$\Omega_{m}h^2$~\cite{eisenstein-1998-496}.

For these reasons, a number of ambitious future surveys have been
proposed aiming to measure the BAO peak at multiple redshifts with
very high accuracy~\cite{Bassett:2005kn,Peacock:2006kj}. Roughly,
one expects to measure $D_A$ from the acoustic scale with an
accuracy~\cite{Peacock:2006kj} $(V/5h^{-3}\,{\rm Gpc})^{-1/2}(k_{\rm
max}/0.2h\,{\rm Mpc}^{-1})^{-1/2}$ where $V$ is the volume of the
survey, and $k_{\rm max}$ is the comoving wavenumber at which the
power spectrum peaks. Currently for the Sloan Digital Sky Survey
(SDSS), the error is of order $4\%$ at $z\sim
0.4$~\cite{Eisenstein:2005su}. These errors are expected to go down
\cite{Aubourpnc:2007} to~ $2.8\%$ at $z\sim 0.4$ by 2008 when SDSS
has accumulated 8000 square degrees; to $1.1\%$ at $z\sim0.6$ and
$1.4\%$ at $z\sim2.5$ (SDSS-III, 10000~deg$^2$); $1\%$ at $z\sim1$
and $1.5\%$ at $z\sim2.8$ (SUBARU+NOAO, 2013); and $1.2\%$ at
$z\sim0.4,0.6,1$ (GWFMOS, 2000~deg$^2$). Amid this excitement, many
studies have been carried out searching for sources of systematic
errors in BAO
measurement~\cite{Guzik:2006bu,Huff:2006gs,Eisenstein:2006nj}
focusing mostly on the non-linear evolution of structure,
scale-dependent bias and errors in the survey window function
estimation, the first two leading to effects smaller than a percent
on $D_A$ while the latter has to be kept below 2\% not to bias the
acoustic scale by more than 1\% at $z=1$. Given the great wealth of
data that will become available, the present analysis focuses mainly
on galaxy surveys and on the resulting galaxy correlation function.
It is necessary to stress, however, that the same kind of analysis
can be applied equally well to the two point correlation function of
different classes of objects, notably the $10^5$ high redshift QSO
that are expected to be surveyed by SDSS-III.

In this paper we show that weak lensing is not necessarily
negligible at the level of accuracy that will be attained by future
surveys. This effect is reminiscent of the weak lensing of Type Ia
supernovae that induces an external dispersion in their observed
brightnesses, comparable in magnitude to the intrinsic dispersion
for redshifts
$z>1$~\cite{Dalal:2002wh,Williams:2004jr,Wang:2004ax,Menard:2004sm}.
In that case, the main effect of lensing is to magnify the
background objects; in this case, while magnification can cause one
effect, distortion of the BAO rings leads to an additional effect.

Light is deflected by large scale structure as it travels from
sources in a deep survey to us, and this lensing leads to two
sources of error in BAO. First, the position we assign to a given
galaxy based on its redshift and angular position on the sky does
{\it not} correspond to its actual position. When we measure the
correlation function by assigning distance to all pairs of galaxies,
we are therefore inevitably making an error. We place a pair of
galaxies in a given distance bin, but they may well belong in a
different bin. This effect tends to smooth out features in the
correlation function so will not change the position of the BAO peak
but will change its height and width. An almost identical effect
acting on the CMB is well-known (see Ref.~\cite{Lewis:2006fu} for a
recent review) and now even included in the codes which compute the
CMB spectrum.

In the case of a galaxy (or QSO) survey, there is a second effect
that does not affect the CMB.  Surveys are generally magnitude
limited, so they include only objects brighter than some threshold.
Not only does lensing change the apparent position of galaxies, but
it also changes their
magnitude~\cite{Mellier:1998pk,Bartelmann:1999yn}: it will brighten
some galaxies, otherwise too faint to be detected, and push them
over threshold and de-magnify others, causing them to drop out of
the survey. This effect, dubbed {\it magnification bias}, is also
well-known especially in studies of the correlation between quasars
and galaxies~\cite{menard-2002-386,Matsubara:2000pr} where it has
been
detected~\cite{Scranton:2005ci,Myers:2005ub,Mountrichas:2007jn}. Its
effect induces a non-zero contribution to the correlation function
even if the underlying galaxy population was completely
random~\cite{Moessner:1997vm,jain-2002-580}.

The paper is organized as follows. In Sec.~\ref{Sec:smooth} we
explicitly compute the effect of the re-mapping due to the lensing
and the smoothing it induces on the correlation function. In
Sec.~\ref{Sec:magbias} we then turn to the analysis of the
magnification bias. In Sec.~\ref{Sec:discussion} we discuss the
impact of these two effects on BAO measurements. Finally, all the
technicalities are gathered in the Appendices.

\section{Smoothing of the Correlation Function}\label{Sec:smooth}

\newcommand\delx{{\bf \delta x}}
\newcommand\delt{{\bf \delta \theta}}
\newcommand\delxs{{\delta x}}

The effect we consider in this section arises because photons travel
on geodesics and these in turn depend on the energy density
distribution present between the source and the observer. This
implies that the observed position of a source galaxy (denoted by
$\vec{x}_o$) and its actual position (denoted by $\vec{x}_s$) will
in general differ. The galaxy number overdensity at a given position
$\vec{x}$ is defined by
\begin{equation}
    \delta^g(\vec{x})\equiv\frac{n(\vec{x})-\bar{n}}{\bar{n}},
\end{equation}
where $n(\vec{x})$ is the galaxy number density at position
$\vec{x}$ and $\bar{n}$ is the mean density that would be measured
if all galaxies were uniformly distributed.

The correlation function is defined as the two point function of the
observed $\delta$ field. The {\it observed, lensed} correlation
function will in general differ from the {\it unlensed} correlation
function. What is actually measured is the correlation between
galaxies overdensities at \textit{observed} positions
\begin{equation}
    \xi_{\rm{obs}}(r)=\xi_{GG}(r)
    \equiv\langle\tilde\delta^g(\vec{x}_o)\tilde\delta^g(\vec{x}'_o)\rangle_{|\vec{x}_o-\vec{x}'_o|=r},
\end{equation}
The unlensed correlation function on the other hand is the
correlation between overdensities at \textit{actual} source
positions $\xi_s(r)$. Since we are interested in large scales, in
what follows we assume a linear bias model (not depending on scales
or redshift~\cite{Simon:2006au}) so that galaxies trace the
underlying dark matter fluctuations and $\delta^g(\vec{x})=b
\delta(\vec{x})$. This assumption in turn implies that galaxies and
matter overdensity correlation functions (resp. denoted by
$\xi_{GG}$ and $\xi$) are simply related by a multiplicative
constant $b^2$: in what follows we focus on the latter even though
the results apply to the former as well. To find the connection
between the lensed and unlensed correlation functions, we use the
fact that weak lensing induces a re-mapping of the density field as
$\tilde\delta(\vec x_o)=\delta(\vec x_s)$ where the two positions
are related by the displacement vector field as $\vec x_o=\vec
x_s+\vec\zeta(\vec x_s)$. A Taylor expansion leads
\begin{equation}
    \tilde\delta(\vec{x}_o)=\delta(\vec{x}_s)
    +\frac{\partial\delta}{\partial x_i}\zeta^i
    +\frac{1}{2}\frac{\partial^2\delta}{\partial x^i\partial x^j}\zeta^i \zeta^j,
\end{equation}
where we have kept terms up to second order in the displacement. It follows that (see
Appendix~\ref{ap:sm} for details)
\begin{eqnarray}
    \xi_{\rm{obs}}(r)&=&\xi_s(r)+\delta\xi_{sm}(r),\\
    \delta\xi_{sm}(r)&=&\left[\langle\zeta^i(\vec x)\zeta^j(\vec x)\rangle-\langle\zeta^i(\vec x)
    \zeta^j(\vec y)\rangle\right]\nonumber\\
    &\times&\left[\frac{\delta^{ij}}{r}\frac{\dd\xi(r)}{\dd r}
    +\frac{r^i
    r^j}{r}\frac{\dd}{\dd r}\left(\frac{\xi'(r)}{r}\right)\right].\label{eq:xiro}
\end{eqnarray}
In this expression the effect of lensing appears only at second
order because $\langle\vec\zeta\rangle=0$.

The additional term $\delta\xi_{sm}$ is due to lensing and tends to
smooth out the correlation function. A simple way to understand it
is to think of the BAO peak as a circle on the sky. When the light
from this circular feature passes through an overdense
line-of-sight, it will appear larger so the feature will move to
larger radius. Conversely, it can appear smaller if the intervening
matter is underdense. Since there is an excess of circles at the
peak scale, most fluctuations will push a circle out of the peak bin
and into an adjacent bin. The net result is that the peak will be
smoothed out: the correlation function will be enhanced below and
above the peak and reduced at the peak. This is encoded in
\ec{xiro}. To further understand this effect, let us consider the
two factors appearing in the lensing term separately. The zero-lag
two-point function $\langle\zeta^i(\vec x)\zeta^j(\vec x)\rangle$ is
simply the variance of the deflection distance while
$\langle\zeta^i(\vec x)\zeta^j(\vec y)\rangle$ is the correlation
between the deflection experienced by two light rays emitted a
distance $r$ from one another. If $r$ is very small, then
$\langle\zeta^i(\vec x)\zeta^j(\vec y)\rangle$ will be very close to
$\langle\zeta^i(\vec x)\zeta^j(\vec x)\rangle$, that is, two
galaxies separated by $r$ will have their geodesics deflected by the
same amount. In that case, the terms in the first set of square
brackets will cancel and there will be no effect: the circles on the
sky will simply be shifted with no distortion in their sizes.
However, if $r$ is large enough then there will be little
correlation between the deflections of galaxies separated by $r$ and
the second term will be smaller than the first. Moving on to the
second set of square brackets, notice that the smoothing term can be
significant where the first or the second derivative are large. This
in turn implies that the departure of $\xi_o(r)$ from $\xi_s(r)$ can
become non-negligible especially in the region of the BAO peak.

%%%%%%%%%%%%%%%%%%%%%%%%%%%%%%%%%%%%%%%%%%%%%%%%%%%%%%%%%%%%%%%%%%
%%%%%%%%%%%%%%%%%%%%%%%%%%%%%%%%%%%%%%%%%%%%%%%%%%%%%%%%%%%%%%%%%%
\begin{figure}
\includegraphics[width=0.98\columnwidth]{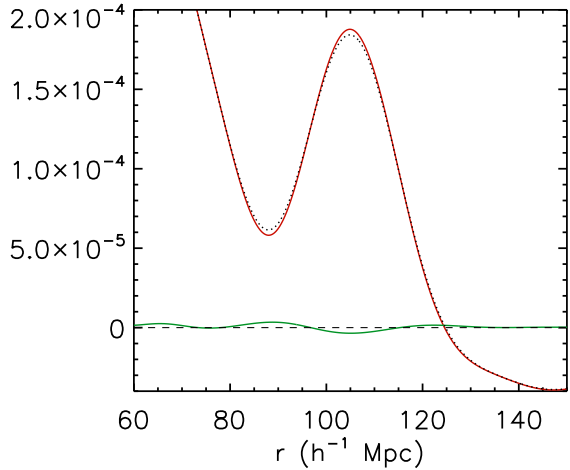}
\caption{\label{Fig:1} Correction due to position re-mapping for
background galaxies at $z=2.5$. Solid red curve is the galaxy-galaxy
correlation function assuming a constant bias factor $b=1$. Green
solid is the re-mapping term. Around $r=100h^{-1}$Mpc it is negative
while on either side it is positive. In other words lensing smooths
out the BAO peak, as can be seen directly from the dotted curve
which includes both contribution. }
\end{figure}
%%%%%%%%%%%%%%%%%%%%%%%%%%%%%%%%%%%%%%%%%%%%%%%%%%%%%%%%%%%%%%%%%%
%%%%%%%%%%%%%%%%%%%%%%%%%%%%%%%%%%%%%%%%%%%%%%%%%%%%%%%%%%%%%%%%%%

\begin{figure}
\includegraphics[width=0.98\columnwidth]{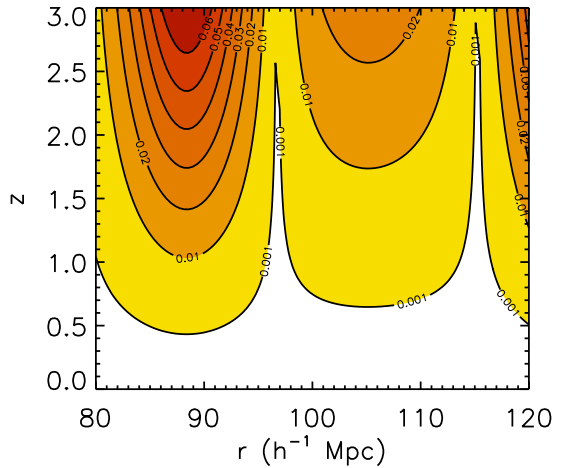}
\caption{\label{Fig:2} Absolute value of the ratio of the
correlations due to position re-mapping and the galaxy-galaxy
correlation function. The lensing contribution is of the order of a
few percent in the region where the correlation function exhibits
the BAO peak.}
\end{figure}

As shown in Appendix~\ref{ap:defl}, adopting the Limber approximation all the
terms appearing in equation (\ref{eq:xiro}) can be recast as integrals over the dark matter
power spectrum. The smoothing correction can then be expressed as
\be
  \delta\xi_{sm}(r)=\xi''\left(\frac{I_1}{2}-I_2+I_3\right)+\frac{\xi'}{r}\left(\frac{I_1}{2}-I_3\right),
  \ee
  where
\begin{eqnarray}
  I_i \equiv \frac{9\Omega_m^2H_0^4}{4}\int\!\!\frac{\dd k}{2\pi k}
  \int_0^{\chi_s}\dd\chi \,W^2(\chi_s,\chi) P_{\delta}(k,\chi)K_i(k,\chi),\eql{kint}\\
  K_i(k,\chi) \equiv\left\{
        \begin{array}{ll}
        1, & i=1 \\
        J_0(k\chi), & i=2 \\
        J_1(k\chi)/(k\chi), \quad & i=3
        \end{array}
   \right.
\end{eqnarray}
and where $W(\chi_s,\chi)=2\chi_s(1-\chi/\chi_s)$ is the window
function, $\chi_s$ is the comoving distance to the source, and a
flat FRW background is assumed throughout. We integrate over the
fully evolving nonlinear power spectrum using the procedure of Smith
et al.~\cite{smith}. The results are depicted in Fig.~\ref{Fig:1},
for simplicity in the case where all galaxies are at the same
redshift. We see that smoothing occurs at the percent level, not a
surprising result given similar conclusions in the case of the CMB.
Nonetheless, this smoothing may need to be accounted for when
analyzing future surveys. This fact is further stressed in
Fig.~\ref{Fig:2}, where the ratio between the smoothing term and the
full correlation function is plotted for the redshift range that
likely will be probed by future surveys.

%%%%%%%%%%%%%%%%%%%%%%%%%%%%%%%%%%%%%%%%%%%%%%%%%%%%%%%%%%%%%%%%%%
%%%%%%%%%%%%%%%%%%%%%%%%%%%%%%%%%%%%%%%%%%%%%%%%%%%%%%%%%%%%%%%%%%
%%%%%%%%%%%%%%%%%%%%%%%%%%%%%%%%%%%%%%%%%%%%%%%%%%%%%%%%%%%%%%%%%%
%%%%%%%%%%%%%%%%%%%%%%%%%%%%%%%%%%%%%%%%%%%%%%%%%%%%%%%%%%%%%%%%%%

There are some important conclusions to be drawn from this analysis.
First, this lensing-induced correlation is unavoidable and has
nothing to do with the way measurements are carried out. This means
that such a contribution will arise even if the ``perfect survey''
with an infinite limiting magnitude is carried out, simply because
geodesics are sensitive to the matter distribution present between
the source and the observer. Second, the previous derivation does
not depend in any way on the class of objects that are being
surveyed: it can be applied equally well to future QSO catalogs.
Third, although the corrections to the correlation function are
larger than a percent, these corrections do \textit{not} alter the
BAO peak position. For example, at $z=2.5$ this lensing-induced
correlation weights about $1.5\%$, but the shift in the peak
position is only $0.01\%$. BAO therefore turn out to be standard
rulers that are very \textit{robust} with respect to this effect.
Finally, the dominant contributions to the lensing-induced terms
arise from modes that are still in the linear regime: we
carried out the integrations using the fully evolving nonlinear
power spectrum, but the results do not change if the linear power
spectrum is used. Physically this is because it is large structures
that are most responsible for deflections; the $k$-integral in
\ec{kint} peaks at the same place the power spectrum does, roughly
$k\simeq0.02h$Mpc$^{-1}$, deep in the linear regime. An important
implication of this is that the correction is very easy to
implement.

\section{Magnification Bias}\label{Sec:magbias}

We now move to consider a second, different effect that can also
impact the determination of the correlation function and the
measurement of the BAO peak position. While the effect analyzed in
the previous section was an unavoidable consequence of any metric
theory of gravity and of the rich structure in the universe, the
effect considered here is related to the fact that actual surveys
include only objects that are brighter than a limiting threshold.

Weak gravitational lensing acts in two ways when such surveys are
carried out. First, it can brighten some objects pushing them over
the threshold and de-magnify others causing them to drop out of the
survey. Second, it can stretch a given patch of sky making it larger
or smaller than it actually is, thus increasing or decreasing the
measured number density. If $n_0(f,z)$ is the number of objects with
flux larger than $f$ per unit solid angle between $z$ and $z+\dd z$
in the absence of lensing, the observed number density in direction
$\hat n$ will be $n_0[f/\mu(\hat n,z),z]/\mu(\hat n,z)$ where
$\mu[(\hat n,z),z]=1/[(1-\kappa)^2-\gamma^2]$ is the magnification
expressed in terms of the convergence $\kappa$ and the shear
amplitude $\gamma$. In the simple case where $\mu$ is assumed
independent of $z$, this implies that $n(m,\hat n)=n_0(m)[\mu(\hat
n)]^{2.5s-1}$ where $s$ is the slope of the number density as a
function of the magnitude $m$, $s\equiv \dd\ln n_0(m)/\dd m$.
Focusing for the time being on the correlation function for
galaxies, it is important to keep in mind that the magnitude and the
sign of this effect (over- or under-density for $\mu>1$) depends on
the slope of the luminosity function, which in turn is different for
different classes of objects and can present redshift evolution
\cite{Scarlata:2006xi}.

Proceeding along the same lines of the analysis carried out by
Moessner and Jain \cite{Moessner:1997vm} on the angular correlation
function, we find that the fluctuation in the galaxy number density
at a given position $\delta(\vec{x})$ receives contributions from
two sources: the true galaxy clustering, denoted by
$\delta^g(\vec{x})$ and the variation in the number density due to
magnification of the galaxy population induced by lensing, denoted
by $\delta^l(\vec{x})$. The observed overdensity is the sum of these
two, $\delta(\vec{x})\equiv\delta^g(\vec{x})+\delta^l(\vec{x})$, so
\begin{eqnarray}\label{1:defxi}
\xi_{\rm{obs}}(r)&=&\langle\delta(\vec{x}) \delta(\vec{y})\rangle
=\langle \delta^g(\vec{x})\delta^{g}(\vec{y})\rangle+ \langle
\delta^g(\vec{x})\delta^{l}(\vec{y})\rangle\nonumber\\
&+&\langle\delta^{l}(\vec{x})\delta^{g}(\vec{y})\rangle+
\langle\delta^{l}(\vec{x})\delta^{l}(\vec{y})\rangle. \eql{xitot}
\end{eqnarray}
where $|\vec{x}-\vec{y}|=r$.

The first term in Eq.~(\ref{1:defxi}) is the true galaxy-galaxy
correlation function. As mentioned in the previous section, in the
linear bias approximation this term is directly proportional to the
matter correlation function.
\begin{equation}
    \langle \delta^g(\vec{x})\delta^{g}(\vec{y})\rangle =
    \xi_{GG}(r)=b^2\xi(r).
\end{equation}

The second and third terms are the contributions to the observed
correlation function arising from correlating a galaxy overdensity
arising because of lensing along the line of sight with a true
matter overdensity. The last term is the contribution to the
correlation function due to two overdensities which are generated
both because of lensing along the line of sight to the source
galaxies.

\begin{figure}
\includegraphics[width=0.98\columnwidth]{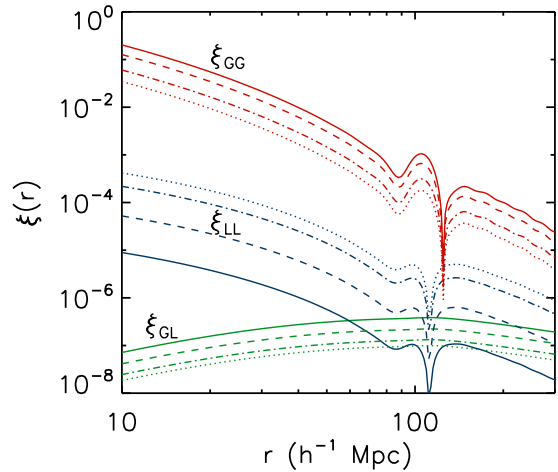}
\caption{\label{Fig:3} Correlation functions at different redshifts:
$\xi_{GG}$ (red), $\xi_{GL}$ (green) and $\xi_{LL}$ (blue). In each
case solid curves corresponds to galaxies at $z=0.5$, dashed curves
$z=1$, dot-dashed curves  $z=2$ and dotted curves $z=3$. For
illustration purposes here we set $b=1$ and $2.5s-1=1$ and we assume
a flat FRW universe with $\Omega_m=0.3$ and $\Omega_{\Lambda}=0.7$.}
\end{figure}

We now consider the two cross terms $\langle
\delta^g(\vec{x})\delta^{l}(\vec{y})\rangle$. These arise from the
correlation of overdensities due to clustering with an apparent
overdensity due to lensing. To proceed further, let us recall the
expression for $\delta^{l}$ originally derived by Broadhurst, Taylor
and Peacock \cite{Broadhurst:1994qu} and also used in Moessner and
Jain~\cite{Moessner:1997vm}
\begin{equation}\label{eq:11}
\delta^{l}(\vec{x})=(5s-2)\kappa(\vec{x}),
\end{equation}
where as before $s$ is the logarithmic slope of the number counts.
The convergence is determined by an integral along the line of sight
\begin{equation}
\kappa(\vec{y})\equiv\frac{3\Omega_m H_0^2}{2}\int_0^{\chi_s(\vec{y})}\dd\chi
 W_L(\chi_s,\chi)\frac{\delta(\vec y,\chi)}{a},
\end{equation}
where $\chi_s$ is the comoving distance to the source galaxy and
$W_L(\chi_s,\chi)\equiv\chi(\chi_s-\chi)/\chi_s$ is the lensing
geometric factor. Using this relation, we derive (see
Appendix~\ref{ap:gl}) that
\begin{equation}
\xi_{GL}(r)\equiv \langle\delta^g\delta^l\rangle = C  b r
\frac{1+\chi_s a(\chi_s) H(\chi_s)}{a(\chi_s)\chi_s}
  I_{gl},
\label{xigl}
\end{equation}
where for sake of brevity we have defined the constant
\begin{equation}
C\equiv\frac{3}{2}(5s-2)\Omega_m H_0^2,
\end{equation}
and the integral
\begin{equation}
I_{gl}\equiv\int\frac{\dd k}{2\pi}P_{\delta}(k,z)J_1(kr),
\end{equation}
where $J_1(x)$ is the Bessel function of the first kind. The factor of $r$ in the prefactor of Eq.~(\ref{xigl})
indicates that not all lensing matter along the line of sight
contributes to this cross term. Rather, only within a distance $r$
are regions with more galaxies likely to produce a larger lensing
signal. If there is a huge excess of matter along the line of sight
but far from the source galaxy, the magnification it produces will
just as likely be associated with a background underdensity as
overdensity, so the correlation $\langle\delta^g\delta^l\rangle$ due
to it will be zero. As a result, this cross term will be largest for
low redshift background galaxies. After all, the matter field is
more clustered at late times, and the cross term is most sensitive
to clustering at the background galaxy redshift. Fig.~\ref{Fig:3}
verifies these two features. The cross term is indeed very small and
does get smaller as one moves to higher redshift.

We now turn to the lensing-lensing term in the correlation function,
which is explicitly given by (see Appendix~\ref{ap:gl})
\begin{equation}
\xi_{LL}(r) \equiv \langle\delta^l\delta^l\rangle=C^2
\int_0^{\chi_s}\dd\chi
 \left[\frac{W_L(\chi_s,\chi)}{a(\chi)}\right]^2 I_{ll}(\chi),
 \label{xill}
\end{equation}
where in this case
\begin{equation}
I_{ll}(\chi)\equiv\int\frac{k\,\dd k}{2\pi} J_0(kr)P_\delta(k,\chi).
\end{equation}
Here there is no prefactor of $r$: all structures along the line of
sight can contribute. As a result, this term is largest when the
background galaxies are at high redshifts and can get lensed by as
much structure as possible.

The result of the calculation of these three contributions is shown
in Fig.~\ref{Fig:3}, where $(2.5s-1)=1$ is assumed and the
integrations are carried out considering the fully evolving
\textit{non-linear} power spectrum obtained applying the procedure
of Smith et al.~\cite{smith}. Four sets of curves are obtained,
corresponding to sources at redshift of 0.5, 1, 2, 3. From
Fig.~\ref{Fig:3}, we see that while the amplitude of $\xi_{GG}$ and
$\xi_{GL}$ decreases for increasing redshift, the amplitude of
$\xi_{LL}$ increases for increasing redshift, thus contributing more
and more to the full correlation function.

%%%%%%%%%%%%%%%%%%%%%%%%%%%%%%%%%%%%%%%%%%%%%%%%%%%%%%%%%%%%%%%%%%
%%%%%%%%%%%%%%%%%%%%%%%%%%%%%%%%%%%%%%%%%%%%%%%%%%%%%%%%%%%%%%%%%%
\begin{figure}
\includegraphics[width=0.98\columnwidth]{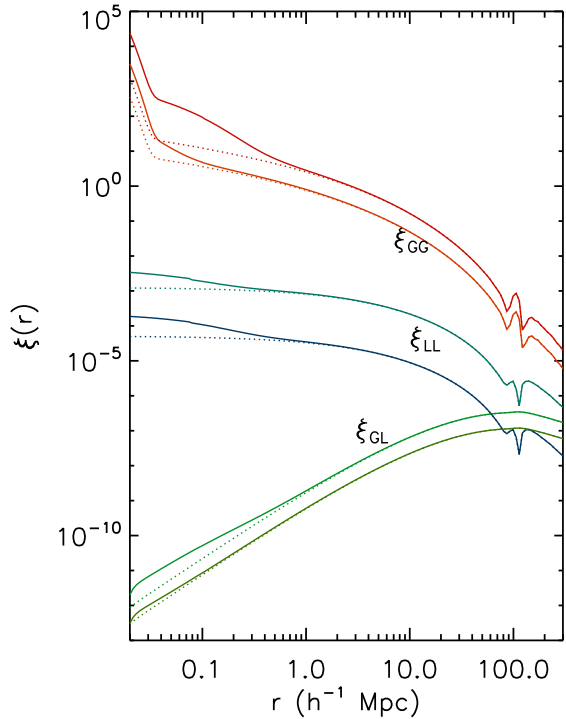}
\caption{\label{Fig:nonlinearities} Calculations of the $\xi_{GG}$,
$\xi_{GL}$ and $\xi_{LL}$ terms carried out for $z=0.5$ (darker
curves) and $z=2.0$ (lighter curves). The solid lines are obtained
integrating over the fully evolving non-linear power spectrum
obtained using the procedure of Smith et al.~\cite{smith}, while the
dotted ones are obtained integrating over the linear power spectrum.
In the range where the correlation function exhibits the BAO peak
non-linearities do not significantly contribute. Again, $b=1$ and
$2.5s-1=1$ are assumed for illustration purposes.}
\end{figure}

%%%%%%%%%%%%%%%%%%%%%%%%%%%%%%%%%%%%%%%%%%%%%%%%%%%%%%%%%%%%%%%%%%
%%%%%%%%%%%%%%%%%%%%%%%%%%%%%%%%%%%%%%%%%%%%%%%%%%%%%%%%%%%%%%%%%%
In Fig.~\ref{Fig:nonlinearities} two sets of curves are shown,
obtained integrating either over the non-linear power spectrum of
Smith et al.~\cite{smith} or over the linear power spectrum. We see
that non-linearities significantly contribute to the correlation
function on scales that are much smaller than the ones where the BAO
peak is located. As in the case of the lensing-induced correlations
considered in the previous section, this result implies that our
conclusions do not depend on the details of the non-linear evolution
of structure.

%%%%%%%%%%%%%%%%%%%%%%%%%%%%%%%%%%%%%%%%%%%%%%%%%%%%%%%%%%%%%%%%%%
%%%%%%%%%%%%%%%%%%%%%%%%%%%%%%%%%%%%%%%%%%%%%%%%%%%%%%%%%%%%%%%%%%
%%%%%%%%%%%%%%%%%%%%%%%%%%%%%%%%%%%%%%%%%%%%%%%%%%%%%%%%%%%%%%%%%%
%%%%%%%%%%%%%%%%%%%%%%%%%%%%%%%%%%%%%%%%%%%%%%%%%%%%%%%%%%%%%%%%%%

%To quantify to what extent the magnification bias effects will be
%important in the determination of the BAO peak, we calculate the
%ratio between the lensing-induced terms $\xi_{GL}+\xi_{LL}$ and the
%full correlation function. This is summarized in Fig.~\ref{Fig:4}.

It is now necessary to stress that equations (\ref{xigl}) and
(\ref{xill}) respectively depend linearly and quadratically on the
factor $(2.5s-1)$. This factor is crucial in determining the weight
of the magnification bias contribution to the correlation function
of the class of objects considered. In particular, despite the fact
that the actual values of $s$ depend on the survey, they are smaller
for galaxies ($s\sim[0.2,0.6]$) \cite{Fort:1996sr,Yasuda:2001xf}
than for QSO ($s\sim[-2,1.6]$) \cite{Scranton:2005ci,Menard:2003vf}.
This means that the $(2.5s-1)$ factors appearing in front of the
magnification bias terms are smaller (larger) than unity for
galaxies (QSO):\footnote{Depending on the class of objects
considered, the curves appearing in Fig.~\ref{Fig:3}
and~\ref{Fig:nonlinearities} need to be rescaled accordingly.} while
the magnification bias is suppressed for the galaxy correlation
function, it can actually play a significant role for the QSO
correlation function. This in turn strengthens the conclusion of the
previous section and points toward a well defined class of objects:
BAO measured through the correlation function of \textit{galaxies}
represent very robust standard rulers, only marginally affected by
lensing even at moderately high redshifts. By the same token, given
the scaling of the magnification bias terms with respect to $C$, the
correlation function of QSO \cite{daAngela:2006mf} can receive
significant contributions from the lensing-lensing term, especially
at high redshift. If one is interested in extracting information
from the lensing-lensing signal, then the QSO correlation function
seems the way to go.

%%%%%%%%%%%%%%%%%%%%%%%%%%%%%%%%%%%%%%%%%%%%%%%%%%%%%%%%%%%%%%%%%%
%%%%%%%%%%%%%%%%%%%%%%%%%%%%%%%%%%%%%%%%%%%%%%%%%%%%%%%%%%%%%%%%%%
%\begin{figure}
%\includegraphics[width=0.98\columnwidth]{pap_xi_zoom.eps}
%\caption{\label{Fig:4} Ratio of the magnification bias terms
%$\xi_{GL}$ and $\xi_{LL}$ to the full correlation
%function. Note that in the region where the
%correlation function exhibits the BAO peak the magnification bias
%weights at most a few percent.}
%\end{figure}
%%%%%%%%%%%%%%%%%%%%%%%%%%%%%%%%%%%%%%%%%%%%%%%%%%%%%%%%%%%%%%%%%%
%%%%%%%%%%%%%%%%%%%%%%%%%%%%%%%%%%%%%%%%%%%%%%%%%%%%%%%%%%%%%%%%%%

\section{Discussion}\label{Sec:discussion}

To understand the role of lensing, recall that
the measurement of the BAO bump appearing in the correlation
function conveys information about two physical scales
\cite{eisenstein-1998-496,Hu:2001bc}. The centroid of the bump is
related to the sound horizon, that is the comoving distance
that a sound wave can travel between the big bang and recombination.
The amplitude of the bump on the other hand is directly related to
the matter-radiation equality period. We have shown that weak
lensing can affect the measurement of baryon acoustic oscillation
through the correlation function in two very different ways.

First, weak lensing smooths out the acoustic peak in the correlation
function. This smoothing is unavoidable, even if the ``perfect
survey'' could eventually be carried out. This lensing-induced term
smooths the BAO peak at the level of a few percent but the peak
location does not change because of it. In other words lensing does
not affect the measurement of the sound horizon. On the other hand,
the measurement of the matter-radiation equality period (and thus of
$\Omega_m h^2$ \cite{Eisenstein:2005su}) depends on the peak
amplitude and it can therefore be affected at the percent level.

Second, weak lensing can also contribute a magnification bias term
to the correlation function that adds to the unlensed term and that
can potentially obscure the prominent feature of interest. This
magnification bias contributes in different ways depending on the
class of objects used to measure the correlation function. In
particular, while it can significantly affect the correlation
function of QSO, magnification bias turns out to be only marginal
for galaxy surveys.

After taking both effects into account, we
conclude that the BAO feature in the galaxy correlation function
is a robust standard ruler. For instance, in the
case in which all surveyed galaxies are at $z=2.5$, the
lensing-induced correlation is of order $1.5\%$, but the location of
the BAO peak shifts by only $0.01\%$. The same cannot be said about
possible QSO correlation function, however, since in that case
magnification bias (notably, the lensing-lensing term) can play a
significant role.

Finally, let us remark that the sensitivity of the results reported
in Sec.~\ref{Sec:smooth} and \ref{Sec:magbias} on the linear bias
parameter $b$ is different for the different effects. While both the
lensing induced smoothing and the magnification bias would be
affected by a scale dependence of the bias factor, the former would
not be sensitive to a change in the overall magnitude of $b$. On the
other hand, if in the future evidence of ``antibias'' will surface
(as suggested by the recent work of Simon et
al.~\cite{Simon:2006au}) this will lead to an enhancement of the
magnification bias. Since $\xi_{GG}$ depends quadratically on $b$
while $\xi_{LL}$ is independent of it, having $b\sim 0.8$ would
imply that the lensing-lensing term weights -- compared to the
galaxy-galaxy term -- almost twice as much as what estimated in
Sec.~\ref{Sec:magbias}.

 \vskip.5cm
 {\bf Acknowledgements:}  We thank Bhuvnesh Jain, Yannick Mellier and Gary Mamon
 for useful conversations. SD is
 supported by the US Department of Energy. AV thanks the Agence
 National de la Recherche for providing financial support. CS
 acknowledges IAP for hospitality.

\bibliography{v7}
\appendix

\section{Re-mapping due to lensing}\label{ap:sm}

Let us compute the new term in the correlation function due to the
difference between the true position of a galaxy and its apparent
position. Weak lensing re-maps the density contrast field, $\delta$,
according to
\begin{equation}
 \tilde\delta(\vec x) = \delta(\vec y)=\delta(\vec x + \vec\zeta),
\end{equation}
where $\vec\zeta$ is the displacement field related to the
deflection angle $\vec\alpha$. For small displacement, it can be
Taylor expanded as
\begin{equation}
 \delta(\vec x + \vec\zeta)=\delta(\vec x) + \zeta^i\delta_{,i}
 +\frac{1}{2}\zeta^i\zeta^j\delta_{,ij}+\ldots,
\end{equation}
where $\delta_{,i}\equiv \partial\delta/\partial x^i$ and $\zeta^i$ are
evaluated in $\vec x$. It follows that the observed correlation
function $\tilde\xi(r)\equiv\langle\tilde\delta(\vec x)
\tilde\delta(\vec y)\rangle$ is given by
\begin{eqnarray}
 \tilde\xi(r) &\simeq&
   \xi(r) + \langle\zeta^i(\vec x)\zeta^j(\vec y)\rangle\langle\delta_{,i}(\vec x)\delta_{,j}(\vec y)\rangle
   \nonumber \\ &&\qquad
   + \langle\zeta^i(\vec x)\zeta^j(\vec x)\rangle\langle\delta_{,ij}(\vec x)\delta(\vec y)\rangle\\
    &\simeq& \xi(r)
   + \langle \zeta^i(\vec x) \zeta^j(\vec y) \rangle {\partial^2\over \partial x^i \partial y^j} \xi(r)
   \nonumber \\ &&\qquad
   + \langle\zeta^i(\vec x)\zeta^j(\vec x)\rangle {\partial^2\over \partial x^i \partial x^j}
   \xi(r).
\end{eqnarray}
Since $r^i\equiv x^i-y^i$, one easily gets that $\partial
\xi(r)/\partial x^i = (r_i/r) \xi'$ where $\xi'\equiv \dd \xi/\dd r$
and then $\partial^2\xi(r)/ \partial x^i \partial y^j= -
\partial^2\xi(r)/\partial x^i \partial x^j$. It follows that
\begin{eqnarray}\label{eq:a5}
 \tilde\xi(r) &\simeq&
   \xi(r)
   + \langle \zeta^i(\vec x)\left[\zeta^j(\vec x) - \zeta^j(\vec y)\right]
   \rangle {\partial^2\xi(r)\over \partial x^i \partial x^j}.
\end{eqnarray}
The second derivative is explicitly given by
\begin{equation}\label{eq:cortemp}
 {\partial^2\xi\over \partial x^i \partial x^j}  = \frac{1}{r}\left[
  \delta_{ij} \xi' +\hat r_i\hat r_j \,r\left( {\xi'\over r} \right)'
 \right]\ ,
\end{equation}
with $\hat r^i=r^i/r$ and $\delta_{ij}$ here is the Kronecker
symbol. Together with Eq.~(\ref{eq:a5}), this leads to the result
in~\ec{xiro}. Note that $\tilde\xi(0)=\xi(0)$, which is a
consequence of isotropy and of the fact that
$\langle\vec\zeta\rangle=0$.

\section{Displacement correlation Function}\label{ap:defl}

The density field $\delta(\vec y)$ is re-mapped to the observed
density field $\tilde\delta(\vec x)$ with $\vec y = \vec x +\vec
\zeta(\vec x)$ by the lensing effects. The displacement field $\vec
\zeta$ is related to the deflection angle $\vec\alpha$ by
\begin{equation}
 \vec\zeta(\vec x) = D_A(\chi)\vec\alpha(\hat n,\chi),
\end{equation}
where $D_A$ is the comoving angular diameter distance, $\chi$ is the
radial comoving radial distance, and $\hat n$ the direction of
observation [so that choosing the coordinate system orientation with
the $\hat{z}$ axis aligned along the line of sight $\vec x =
(D_A(\chi)\hat n,\chi)$]. This means that we assume that the effect
of gravitational lensing is to shift the position of the galaxies
perpendicularly to the line of sight and that we are neglecting the
(small) effect on the redshift and thus on the radial coordinate.

The deflection angle is related to the gravitational potential
$\Phi$ integrated along the line of sight by
\begin{equation}
 \vec\alpha(\hat n,\chi) = 2\int_0^\chi \frac{D_A(\chi-\chi')}{D_A(\chi)}\nabla_\perp
 \Phi[D_A(\chi')\hat n,\chi']\dd\chi',
\end{equation}
where $\nabla_\perp$ is the gradient in the 2-dimensional plane perpendicular to the line of
sight~\cite{Dodelson,PUbook}. It follows that
\begin{equation}\label{eq:zeta}
 \vec\zeta(\hat n,\chi) = \int_0^\chi W(\chi,\chi') \nabla_\perp
 \Phi[D_A(\chi')\hat n,\chi']\dd\chi',
\end{equation}
with the window function
\begin{equation}\eql{lensker}
 W(\chi,\chi')\equiv 2D_A(\chi-\chi'),
\end{equation}
which reduces for a universe with Euclidean spatial sections to
$W(\chi,\chi') = 2\chi(1-\chi'/\chi)$.

The gravitational potential is expanded in harmonic space with the Fourier convention
\begin{equation}~\label{eq:bracket}
 \Phi(\vec x,\eta) = \int \frac{\dd^3\vec k}{(2\pi)^{3/2}}\Phi(\vec k,\eta)
 e^{i\vec k.\vec x},
\end{equation}
and we define the power spectrum
\begin{equation}
 \langle\Phi(\vec k,\eta)\Phi^*(\vec k',\eta')\rangle =
 P_\Phi(k;\eta,\eta')\delta^{(3)}(\vec k - \vec k'),
\end{equation}
where $\eta$ is the conformal time so that the unperturbed geodesic
equation is $\chi=\eta_0-\eta$.

Using Eq.~(\ref{eq:zeta}) and the Fourier decomposition, the
displacement correlation function is given by
\begin{widetext}
\begin{equation}
 \langle\zeta^i(\vec x)\zeta^j(\vec y)\rangle = \int_0^{\chi_x} W(\chi_x,\chi_1)\dd\chi_1
 \int_0^{\chi_y}W(\chi_y,\chi_2)\dd\chi_2
  \int\!\!\!\!\frac{\dd^3\vec k_1}{(2\pi)^{3/2}}
 \frac{\dd^3\vec
 k_2}{(2\pi)^{3/2}}\,
 \nabla_{\perp1}^i\nabla_{\perp2}^j
 \langle\Phi(\vec k_1,\chi_1)\Phi(\vec k_2,\chi_2)\rangle\,
 \hbox{e}^{i(\vec k_1\cdot\vec X - \vec k_2\cdot\vec Y)},
\end{equation}
\end{widetext}
where $\vec X$ and $\vec Y$ are the equation of the null geodesics
relating the galaxies respectively in $\vec x$ and $\vec y$ to the
observer. We do not state here whether $\vec x=\vec y$ or not. In
spherical coordinates of the position $\vec x$ (resp. $\vec y$) is
$(\chi_x,\hat n_x)$, $\hat n_x$ being a spacelike unit vector and
$\vec X$ (resp. $\vec Y$) given $(\chi_1,\hat n_x)$ with
$\chi_1=\eta_0-\eta$.

We use Eq.~(\ref{eq:bracket}) to express the field correlator and
integrate over $\vec k_2$ to get
%\begin{widetext}
\begin{eqnarray}
 \langle\zeta^i(\vec x)\zeta^j(\vec y)\rangle &=& \int_0^{\chi_x} W(\chi_x,\chi_1)\dd\chi_1
 \int_0^{\chi_y} W(\chi_y,\chi_2)\dd\chi_2\nonumber\\
 &\times&\int\frac{\dd^3\vec k}{(2\pi)^3}
 k^i_\perp k^j_\perp P_\Phi(k;\chi_1,\chi_2)
 \nonumber\\
 &\times&\hbox{e}^{i\vec k_\perp\cdot(\vec X_\perp - \vec Y_\perp)}\hbox{e}^{i
 k_\parallel(\chi_1-\chi_2)},
\end{eqnarray}
%\end{widetext}
where we have decomposed the wave-vector as $\vec k = (\vec
k_\perp,k_\parallel)$. On small scales ($\theta\ll1$), we can make
the Limber approximation~\cite{Bartelmann:1999yn,PUbook} which
exploits the fact that the main contribution to the signal comes
from transverse modes so that $P_\Phi(k;\chi,\chi')\simeq
P_\Phi(k_\perp;\chi,\chi')$. This allows us to integrate over
$k_\parallel$ to get $2\pi\delta^{(1)}(\chi_1-\chi_2)$ and then on
$\chi_2$ to get
\begin{eqnarray}
 \langle\zeta^i(\vec x)\zeta^j(\vec y)\rangle &=& \int_0^{\chi_M} W(\chi_x,\chi')W(\chi_y,\chi')
 \dd\chi'\\
 &&\int\frac{\dd^2\vec k_\perp}{(2\pi)^2}
 k^i_\perp k^j_\perp P_\Phi(k_\perp;\chi')
 \hbox{e}^{i\vec k_\perp\cdot\vec R_\perp},\nonumber
\end{eqnarray}
with $\vec R_\perp\equiv\vec X_\perp - \vec Y_\perp$ and where
$\chi_M={\rm max}[\chi_x,\chi_y]$. In order to evaluate
Eq.~(\ref{eq:cortemp}), we need to compute both $\langle\zeta^i(\vec
x)\zeta_i(\vec y)\rangle$ and $\langle\hat r_i\zeta^i(\vec x)\hat
r_j\zeta^j(\vec y)\rangle$, paying attention to the fact that $\vec
x$ may or may not be equal to $\vec y$.

Let us first concentrate on $\langle\zeta^i(\vec x)\zeta_i(\vec
y)\rangle$. We choose the orientation in the plane perpendicular to
the line of sight such that $\vec k_\perp\cdot\vec
R_\perp=k_\perp\chi\cos\varphi$. If $\vec x=\vec y$, the integration
over $\varphi$ just gives $2\pi$. If $\vec x\not=\vec y$, we have to
integrate $\exp(i k_\perp\chi\cos\varphi)$, which can be easily seen
to give $2\pi J_0(k_\perp\chi)$ by using the series expansion of the
complex exponential in terms of Bessel functions,
\begin{equation}\label{eq:exp}
 \hbox{e}^{ix\cos\varphi}= J_0(x) + 2\sum_{n=1}^\infty i^n\cos(n\varphi)J_n(x).
\end{equation}
We conclude that
\begin{eqnarray}
 \langle\zeta^i(\vec x)\zeta_i(\vec y)\rangle &=&
 \frac{1}{2\pi}\int_0^{\chi_M} W(\chi_x,\chi)W(\chi_y,\chi)\dd\chi
 \\
 &\times&\int
 \left\lbrace
 \begin{array}{l}
 1 \\ J_0(k\chi)
 \end{array}
 \right\rbrace
 P_\Phi(k;\chi,\chi)k^3 \dd k,
 \textrm{ for }
 \begin{array}{l}
 \vec x=\vec y \\ \vec x\not=\vec y
 \end{array}
 .\nonumber
\end{eqnarray}
Let us now turn to the second term $\langle\zeta^i(\vec
x)\zeta^j(\vec y)\rangle\hat r_i\hat r_j$. Again, we choose the
orientation in the plane perpendicular to the line of sight such
that $\vec k_\perp\cdot\vec R_\perp=k_\perp\chi\cos\varphi$. We have
term $k_\perp^2$ is now replaced by $(\vec k_\perp^i\cdot\hat
r_i)^2=k_\perp^2\cos^2\varphi$. If $\vec x=\vec y$, the integration
over $\varphi$ of $\cos^2\varphi$ just gives $\pi$. If $\vec
x\not=\vec y$, we have to integrate $\cos^2\varphi\exp(i
k_\perp\chi\cos\varphi)$. Again, we use the expansion~(\ref{eq:exp})
and only two terms contribute to the integration over $\varphi$:
$J_0(k\chi)\cos^2\varphi$ gives $\pi J_0(k\chi)$ and for $n=2$,
$2i^2\cos 2\varphi \cos^2\varphi$ gives $-\pi J_2(k\chi)$. We
conclude that
\begin{widetext}
\begin{eqnarray}
 \hat r_i\hat r_j\langle\zeta^i(\vec x)\zeta^j(\vec y)\rangle &=&
 \frac{1}{4\pi}\int_0^{\chi_M} W(\chi_x,\chi)W(\chi_y,\chi)\dd\chi
 \int
 \left\lbrace
 \begin{array}{l}
 1 \\ J_0(k\chi)-J_2(k\chi)
 \end{array}
 \right\rbrace
 P_\Phi(k;\chi,\chi)k^3 \dd k,
 \textrm{ for }
 \begin{array}{l}
 \vec x=\vec y \\ \vec x\not=\vec y
 \end{array}
 .
\end{eqnarray}
\end{widetext}
It follows that Eq.~(\ref{eq:cortemp}) reduces to
\begin{equation}\label{eq:cor2}
 \tilde\xi(r)=\xi(r)+A\frac{1}{r}\xi' +Br\left( {\xi'\over r}
 \right)',
\end{equation}
with
%\begin{widetext}
\begin{eqnarray}
 A&=&
 \frac{1}{2\pi}\int k^3\dd k\int_0^{\chi_M} \dd\chi\, P_\Phi(k;\chi,\chi)W(\chi_x,\chi)
 \nonumber\\
 &\times&\left[W(\chi_x,\chi)-W(\chi_y,\chi)J_0(k\chi)\right],
 \\
 B&=&
 \frac{1}{4\pi}\int k^3\dd k\int_0^{\chi_M}
 \dd\chi\,P_\Phi(k;\chi,\chi)W(\chi_x,\chi)\nonumber\\
 &\times&\left[W(\chi_x,\chi)-W(\chi_y,\chi)(J_0(k\chi)-J_2(k\chi))\right].
\end{eqnarray}
%\end{widetext}
To finish, one should give an expression for $P_\Phi$. The Poisson
equation on sub-Hubble scales, $\Delta\Phi = 4\pi G \rho a^2 \delta
= 3\Omega_{m0}H_0^2\delta/a$ implies that
\begin{eqnarray}
 P_\Phi(k;\chi,\chi') &=& P_\delta(k;\eta,\eta') \left(\frac{3\Omega_m H_0^2}{2k^2}\right)^2
 \nonumber\\
 &=& P_\delta(k;0) \left(\frac{3\Omega_m H_0^2}{2k^2}\right)^2\!\!\frac{g(\chi)g(\chi')}{g(0)^2},
\end{eqnarray}
where $g=D/a$ with $D$ being the growth factor of the density
perturbations, $\delta(\vec k,\eta)=D(\eta)\delta_{\rm init}(\vec
k)$, and $P_\delta(k;0)\equiv P_\delta(k)$ the density contrast
power spectrum at redshift 0.

\section{Galaxy-Lensing and Lensing-Lensing Correlations}\label{ap:gl}

In this appendix, we give explicit expressions for the correlators
needed for the computation of the magnification bias.

\subsection{Galaxy-lensing term: $\langle \delta n_1^g\delta n_2^{l *}\rangle$}

First, let us consider the galaxy-lensing cross correlation
function. It corresponds to the second [and third] term in
\ec{xitot}. Going back to the second term of the sum we have
\begin{equation}
\langle \delta n_1^g(\vec{x_0})\delta n_2^{l *}(\vec{y_0})\rangle
  =(5s-2) \langle\delta
  n_1^g(\vec{x_0})\kappa^*(\vec{y_0})\rangle,
\end{equation}
where $\delta n$ is the fluctuation of the galaxy number density and
$\delta n^l$ has been related to the convergence $\kappa$ by
Eq.~(\ref{eq:11}). The l.h.s correlator is given by

\begin{eqnarray}
\langle\delta
  n_1^g(\vec{x_0})\kappa^*(\vec{y_0})\rangle
   &=&\frac{3}{2}\Omega_m H_0^2b
       \int_0^{\chi_L(\vec{y_0})}
       \!\!\!\!\!\!\dd\chi\frac{W_L}{a}
            \langle\delta(\vec{x_0})\delta^*(\vec{y}(\chi))\rangle
   \nonumber\\
  &=& \frac{3}{2}\Omega_m H_0^2b
  \int_0^{\chi_L(\vec{y_0})}\!\!\!\!\!\!\dd\chi\frac{W_L}{a}
\xi[r(\chi)], \label{corrgalaxylensing}
\end{eqnarray}
where $\vec r(\chi)=|\vec{x_0}-\vec{y}(\chi)|$ so that $r(\chi)$ the
distance between the source at $\vec{x}_0$ and the point
$\vec{y}(\chi)$ which is running along the line of sight between the
observer and the source point at $\vec{y}_0$. $W_L$ is the lensing
geometric factor,
\begin{equation}
    W_L(\chi_L,\chi)\equiv \chi(\chi_L-\chi)/\chi_L,
\end{equation}
and $\xi(r)$ is the unlensed (or ``true'') matter correlation
function
\begin{equation}\label{xiofchi}
  \xi[r(\chi)]=\int \frac{k^2dk}{2\pi^2}j_0[kr(\chi)]P(k)=\hat{\xi}(\chi).
\end{equation}
Equation (\ref{corrgalaxylensing}) tells us that the correlation between a true matter overdensity
and a lensed one is given by an integral over the line of sight of the correlation function
relative to the true overdensity and the point running along the line of sight weighted by the
usual lensing factor. We can actually work on the above expression much further in the spirit of
Limber's approximation. Letting for brevity $\langle\delta n_G\delta n_L\rangle=\langle \delta
n_1^g(\vec{x_0})\delta n_2^{l
  *}(\vec{y_0})\rangle$
and $C=(5s-2)3\Omega_m H_0^2/2$ and remembering the dependence on
the sources' positions of the two terms we have
\begin{eqnarray}
 \langle\delta n_G\delta n_L\rangle
    &=& C b
    \int_0^{\chi_L(\vec{y}_0)}\!\!\!\!\!\!\dd\chi \frac{W_L}{a}
    \int \!\!\!\frac{\dd^3\vec k}{(2\pi)^3}\hbox{e}^{i\vec{k}\cdot\vec{r}(\chi)}
    P_{\delta}(k,z)\nonumber\\
  &=&C b\int_0^{\chi_L(\vec{y}_0)}\!\!\!\!\!\!\dd\chi
     \int \!\!\!\frac{\dd^3k}{(2\pi)^3}P_{\delta}(k,z)\frac{W_L}{a}\nonumber\\
   &\times&
     \hbox{e}^{i[k_1\chi_G\theta_1+k_2\chi_G\theta_2+k_3(\chi_G-\chi)]},
\end{eqnarray}
where in the last step we have picked the coordinate system so that $\vec{y}_0=(0,0,\chi_L)$,
$\vec{x}_0=\chi_G(\theta_1, \theta_2, 1)$ and the $\vec{k}$ coordinate system so that $k_3$ is
aligned along $\vec{y}_0$.\footnote{The result will not change if instead we pick
$\vec{y}_0=\chi_L(\theta_1,\theta_2,1)$ and $x_0=\chi_G(0, 0, 1)$.}

So far we have been considering the completely general case in which
$\chi_G\neq\chi_L$. For simplicity let us now set
$\chi_L=\chi_G=\chi_s$. Then
\begin{eqnarray}
  \langle\delta n_G\delta n_L\rangle &=&
     C b\int_0^{\chi_s(\vec{y}_0)}\!\!\!\!\!\!\dd\chi
     \int \!\!\!\frac{\dd^3\vec k}{(2\pi)^3}P_{\delta}(k,z
     )\frac{W_L}{a(\chi)}\nonumber\\
  &\times& \hbox{e}^{i[k_1\chi_s\theta_1+k_2\chi_s\theta_2+k_3(\chi_s-\chi)]}.
\end{eqnarray}
Using the fact that
\begin{equation}
    W_L\exp[ik_3(\chi_s-\chi)] =
-i(\chi/\chi_s)\partial_{k_3} \exp[ik_3(\chi_s-\chi)],
\end{equation}
and then integrating by parts over $k_3$ leads to
\begin{eqnarray}
  \langle\delta n_G\delta n_L\rangle
  &=&-C b\int_0^{\chi_s(\vec{y}_0)}\!\!\!\!\!\!\!\!\!\!\!\dd\chi\int \frac{\dd^3
  \vec k}{(2\pi)^3}\frac{\chi}{\chi_s}\frac{1}{ak}
  \frac{\dd P_{\delta}(k,z)}{\dd k}\nonumber\\
  &\times& \partial_\chi\hbox{e}^{i[k_1\chi_s\theta_1+k_2\chi_s\theta_2+k_3(\chi_s-\chi)]},
\end{eqnarray}
where we have used the fact that for any function of the modulus
only $\partial_{k_3}f(k)=(k_3/k)\dd f(k)/\dd k =
(k_3/k_\perp)\partial_{k_\perp}f(k)$ with $k^2=k_\perp^2+k_3^2$ and
where, again, we have used that $k_3\exp[ik_3(\chi_s-\chi)]=
i\partial_\chi\exp[ik_3(\chi_s-\chi)]$. Now in the Limber's limit in
which the $k_3$ dependence of the power spectrum is neglected, we
can integrate over $k_3$ to get $2\pi\delta^{(1)'}(\chi_s-\chi)$
(where $\delta^{(1)'}$ is the derivative of the Dirac distribution).
Integrating over $\chi$ then leads to
\begin{eqnarray}
  \langle\delta n_G\delta n_L\rangle
      &=&
   \frac{C b}{a_s\chi_s}\int \frac{\dd^2\vec k_\perp}{(2\pi)^2k_\perp}
  \left(1+a_s\chi_s H_s\right)\nonumber\\
  &\times&\partial_{k_\perp} P_{\delta}(k_\perp,\chi_s)
  \hbox{e}^{i[k_1\chi_s\theta_1+k_2\chi_s\theta_2]},
\end{eqnarray}
with $a_s=a(\chi_s)$ and where we have used the fact that $\dd
a/\dd\chi_s=-a^2_s H_s$ with $H_s=H(a_s)$. We integrate over the
angle with the use of Eq.~(\ref{eq:exp}) to get $(2\pi)J_0(r)$ with
$r=\chi_s\theta$ and then make an integration by part to get
\begin{eqnarray}
  \langle\delta n_G\delta n_L\rangle
  &=& C b \left(\frac{1}{a_s\chi_s}+H_s\right)r \nonumber\\
  &\times&\int\frac{\dd k}{2\pi}P_{\delta}(k,z)J_1(kr).
\end{eqnarray}
This leads to the result in Eq.~(\ref{xigl}).

\subsection{Lensing-lensing term: $\langle\delta n_1^{l}\delta n_2^{l *}\rangle$}

The lensing-lensing term
\begin{eqnarray}
\langle\delta n_1^{l}(\vec{x_0})\delta n_2^{l*}(\vec{y_0})\rangle
&=& (5s-2)^2\langle\kappa(\vec{x}_0)\kappa^*(\vec{y}_0)\rangle,
\end{eqnarray}
is directly related to the convergence correlation function and can be computed in a similar way
as the galaxy-lensing term. We start from the expression
\begin{widetext}
\begin{eqnarray}\label{eq:kappa}
\langle\kappa(\vec{x}_0)\kappa^*(\vec{y}_0)
   &=& \frac{9 \Omega_m^2 H_0^4}{4}\int_0^{\chi_1(\vec{x}_0)}d\chi_1
       \int_0^{\chi_2(\vec{y}_0)}d\chi_2
       \frac{\chi_1\left(\chi_1(\vec{x}_0)-\chi_1\right)}{\chi_1(x_0)a(\chi_1)}
\frac{\chi_2\left(\chi_2(\vec{y}_0)-\chi_2\right)}{\chi_2(\vec{y}_0)a(\chi_2)}
\langle\delta(\vec{x}(\chi_1))\delta(\vec{y}(\chi_2))\rangle.
\end{eqnarray}
\end{widetext}
Indeed, when the two sources lie on the same line of sight, it
reduces to the variance of the convergence. However, in general the
two lines of sight -- parameterized by $\vec{x}(\chi_1)$ and
$\vec{y}(\chi_2)$ -- are not identical. In this case, defining
$r'\equiv |\vec{x}(\chi_1)-\vec{y}(\chi_2)|$, it is possible to
recast Eq.~(\ref{eq:kappa}) in terms of the correlation function
$\xi(r')$.

We use the same conventions and notations as in the previous paragraph. We choose the coordinate
system such that $\vec{y}(\chi_2)=(0,0,\chi_2)$ and $\vec{x}(\chi_1)=\chi_1(\theta_1,\theta_2,1)$
and we orient the $k$-space coordinate system so that $k_3$ is parallel to $\vec{y}(\chi_2)$. It
follows that Eq.~(\ref{eq:kappa}) takes the form
\begin{eqnarray}
 \langle\delta n_L \delta n_L\rangle &=&
     C^2\int_0^{\chi_s}d\chi_1 \int_0^{\chi_s}d\chi_2
    \frac{W_L(\chi_1)}{a(\chi_1)} \frac{W_L(\chi_2)}{a(\chi_2)}\nonumber\\
    &&\!\!\!\!\!\!\!\!\!\!\!\!\!\!\!\!\!\!\!\!\!\!\!\!\!\!\!\times
    \int\frac{\dd^3\vec k}{(2\pi)^2}P_\delta(k,z)
    \hbox{e}^{i[k_1\chi_1\theta_1+k_2\chi_1\theta_2+k_3(\chi_1-\chi_2)]}.
\end{eqnarray}
Again, we use the Limber's approximation so that $P_\delta(k,z)\simeq P_\delta(k_{\perp},z)$. In
this limit, the integral over $k_3$ simply gives $(2\pi)\delta^{(1)}(\chi_1-\chi_2)$ and
\begin{eqnarray}
 \langle\delta n_L \delta n_L\rangle &=&
     C^2 \int_0^{\chi_s}\!\!\!\!\dd\chi
   \frac{W_L^2(\chi)}{a^2(\chi)}
   \int\frac{\dd^2\vec k_\perp}{(2\pi)^2}P_\delta(k,\chi)\nonumber\\
   &\times&\hbox{e}^{i\chi(k_1\theta_1+k_2\theta_2)}.
\end{eqnarray}
To finish, we integrate over the angle in the $\vec k_\perp$-plane
to get
\begin{equation}
 \langle\delta n_L \delta n_L\rangle
   = C^2 \int_0^{\chi_s}\dd\chi
   \left[\frac{W_L(\chi)}{a(\chi)}\right]^2I_{LL}(\chi),
\end{equation}
with
\begin{equation}
   I_{ll}(\chi)=\int\frac{k\,\dd k}{2\pi}
   J_0(kr)P_\delta(k,\chi),
\end{equation}
which is exactly the result given in Eq.~(\ref{xill}).

\end{document}